\documentclass[aip,apl,amsmath,amssymb,reprint,]{revtex4-1}
\usepackage{graphicx}
\usepackage{dcolumn}
\usepackage{bm}

\begin{document}

\title{Lithium Niobate Metasurfaces}

\author{Bofeng Gao}
\affiliation{The Key Laboratory of Weak-Light Nonlinear Photonics, Ministry of Education, School of Physics and TEDA Applied Physics Institute, Nankai University, Tianjin 300071, P.R. China}

\author{Mengxin Ren}
\email{ren$_$mengxin@nankai.edu.cn}
\affiliation{The Key Laboratory of Weak-Light Nonlinear Photonics, Ministry of Education, School of Physics and TEDA Applied Physics Institute, Nankai University, Tianjin 300071, P.R. China}

\author{Wei Wu}
\affiliation{The Key Laboratory of Weak-Light Nonlinear Photonics, Ministry of Education, School of Physics and TEDA Applied Physics Institute, Nankai University, Tianjin 300071, P.R. China}

\author{Wei Cai}
\affiliation{The Key Laboratory of Weak-Light Nonlinear Photonics, Ministry of Education, School of Physics and TEDA Applied Physics Institute, Nankai University, Tianjin 300071, P.R. China}

\author{Hui Hu}
\affiliation{School of Physics and Microelectronics, Shandong University, Jinan, Shandong 250100, P.R. China}

\author{Jingjun Xu}
\email{jjxu@nankai.edu.cn}
\affiliation{The Key Laboratory of Weak-Light Nonlinear Photonics, Ministry of Education, School of Physics and TEDA Applied Physics Institute, Nankai University, Tianjin 300071, P.R. China}

\begin{abstract}
Lithium niobate is a multi-functional material, which has been regarded as one of the most promising platform for the multi-purpose optical components and photonic circuits. Targeting at the miniature optical components and systems, lithium niobate microstructures with feature sizes of several to hundreds of micrometers have been demonstrated, such as waveguides, photonic crystals, micro-cavities, and modulators, \textit{et al}. In this paper, we presented subwavelength nanograting metasurfaces fabricated in a crystalline lithium niobate film, which hold the possibilities towards further shrinking the footprint of the photonic devices with new optical functionalities. Due to the collective lattice interactions between isolated ridge resonances, distinct transmission spectral resonances were observed, which could be tunable by varying the structural parameters. Furthermore, our metasurfaces are capable to show high efficiency transmission structural colors as a result of structural resonances and intrinsic high transparency of lithium niobate in visible spectral range. Our results would pave the way for the new types of ultracompact photonic devices based on lithium niobate.
\end{abstract}

\maketitle

Lithium niobate (LiNbO$_3$, LN) has been regarded as a key material in photonics community because of its commercial availability and multi-functional properties,\cite{chen2012,poberaj2012} including excellent visible and infrared optical transparency, large electro-optic, piezoelectric and second order harmonic coefficients, as well as photorefactive capability \textit{et al}.\cite{weis1985} Their electrical and optical properties could be effectively adjusted by ion doping, for example increasing optical damage resistance by magnesium or zirconium dopant.\cite{kong2007,kong2012} Just like silicon as the backbone of electronic industry, LN acts as a multi-purpose material platform for photonics, and opto-electronic systems.\cite{arizmendi2004} During the past decades, most of the attention has been drawn to realize high-density integrated photonic/ optical circuits.\cite{wooten2000,boes2018} And various techniques have been developed to make photonic microstructures using LN, for example thermal diffusion or proton exchange,\cite{kip1998} laser writing,\cite{lin2015} focussed ion milling,\cite{si2011} ion etching,\cite{chen2009} and so on. Until now, the waveguides,\cite{courjal2011,bazzan2015} photonic crystals,\cite{bernal2006,diziain2008,sulser2009,geiss2010,cai2014} micro-cavities,\cite{diziain2015,wang2015,wang2018} optical frequency converters,\cite{diziain2013,geiss2015,wang2017} superbroad-bandwidth modulators\cite{lu2014,wangcheng2018} with the feature sizes of several to hundreds of micrometers have been fabricated. However, how to further reduce the sizes of the LN based photonic elements and meanwhile extend their functionalities are still in pursuit for novel photonic systems with ultra-compact fashion.

\begin{figure} [!b]
\includegraphics[width=85mm]{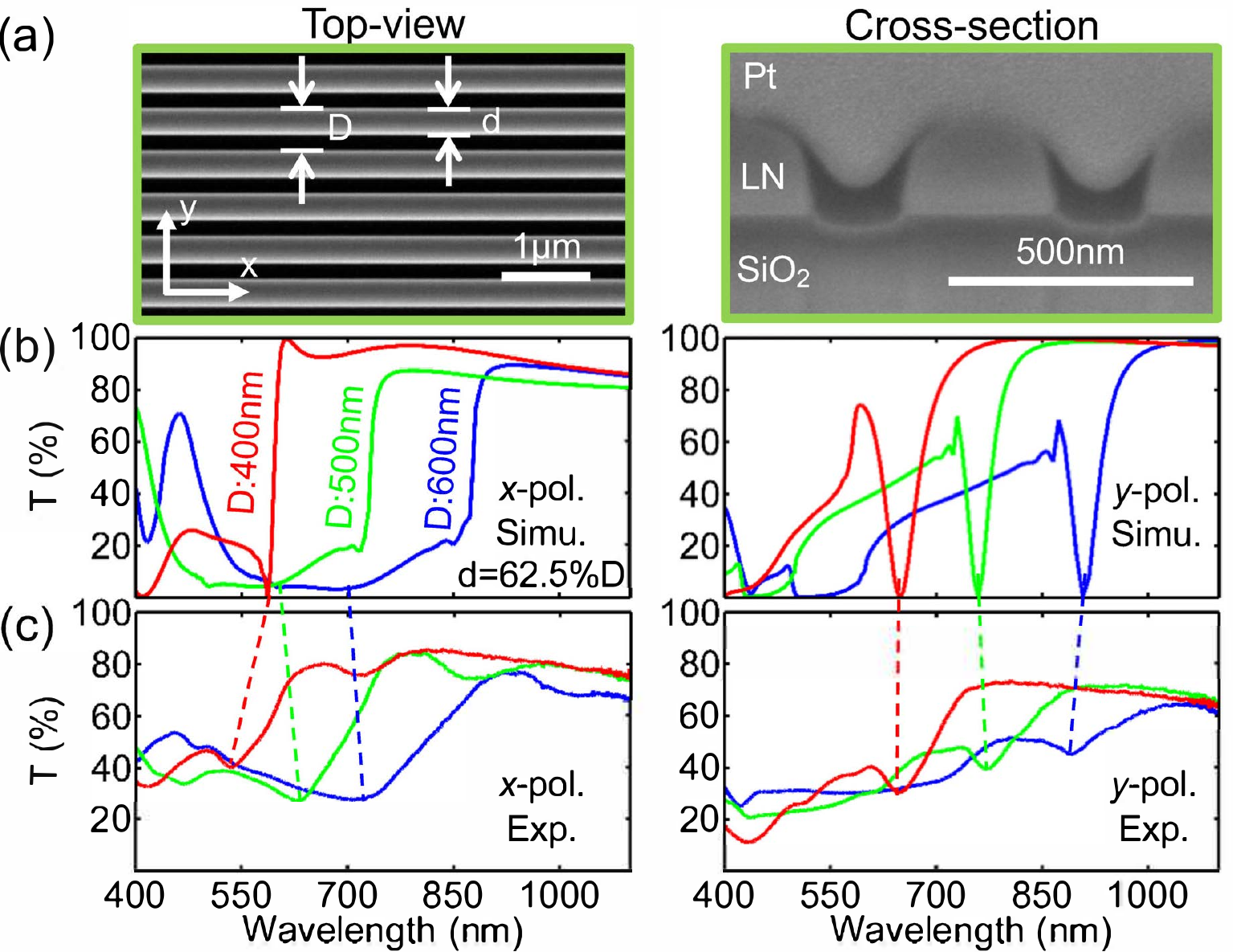} \caption{\label{fig1}
\textbf{The SEM images and spectral properties of LN nanograting metasurfaces.} \textbf{a}, The typical SEM images of fabricated array with $D$=500~nm and $d$=312.5~nm. The cross-section is shown on the right, in which Pt is used as a protection layer for cross-section cutting. The profile of LN is round shaped rather than the designed rectangle. The scale bars are 1000~nm and 500~nm, respectively. \textbf{b,c} The simulated and experimental transmission spectra for arrays with $D$ of 400~nm, 500~nm, and 600~nm for orthogonal polarizations. The geometric parameters used in simulations here are reproduced from the cross-section.}
\end{figure}

\begin{figure*} [!t]
\includegraphics[width=110mm]{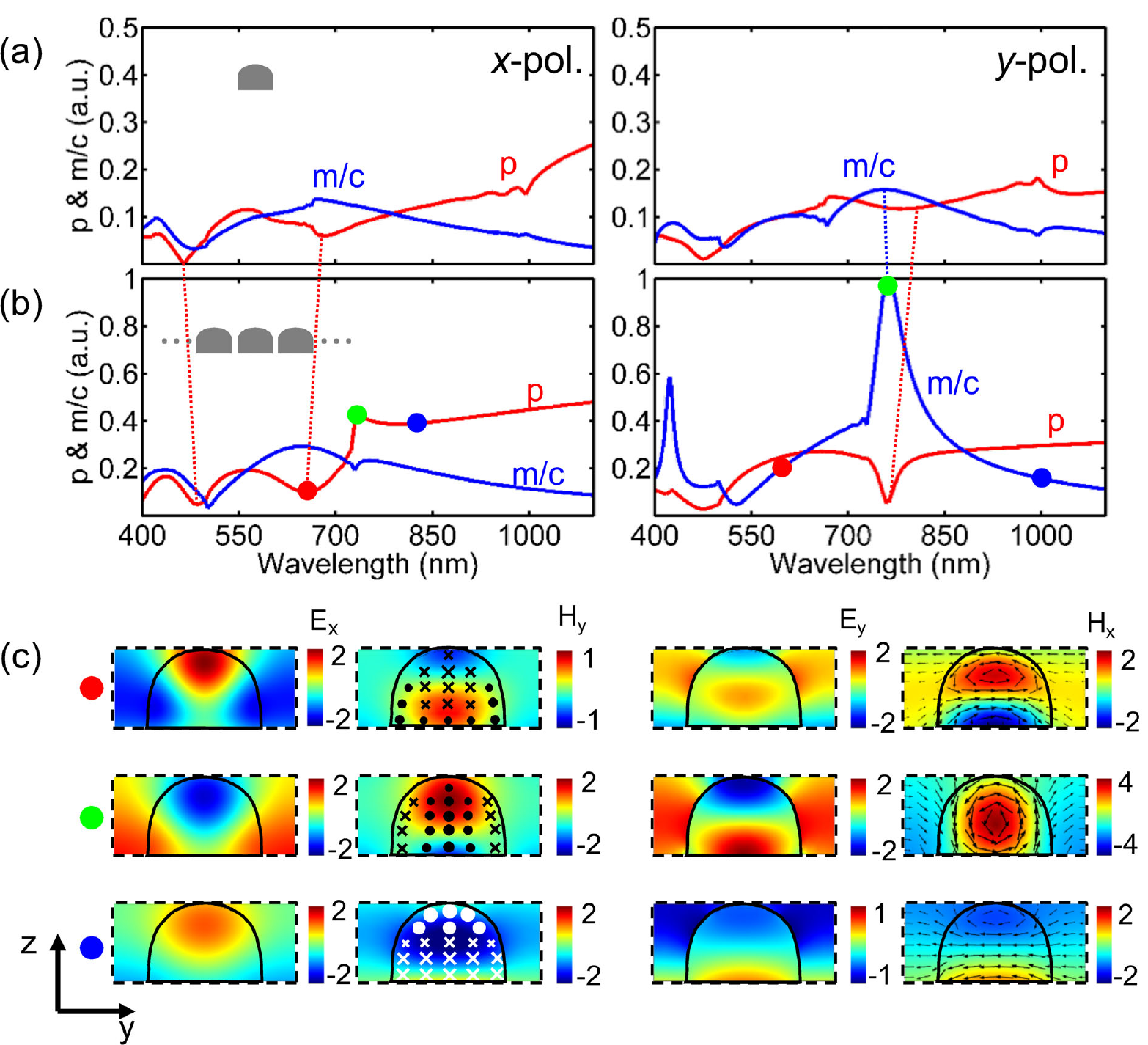} \caption{\label{fig2}
\textbf{The electric (and magnetic) dipole moments induced inside LN and the simulated field maps.} \textbf{a}, The spectra of the electric (\textbf{p}, red curves) and the magnetic (\textbf{m}, blue curves) dipole moments' magnitude induced inside an isolate single LN ridge (shown in the inset) for $x$- and $y$-polarized incident planar waves, respectively. The \textbf{m} is divided by light speed $c$ considering that the emission of electric dipole is $c$ times larger than the magnetic counterpart in vacuum. \textbf{b}, The results for the periodic arrays with $D$=500~nm. \textbf{c}, The field maps at wavelengths marked in (b) on the cross-section for one structure period. The area enclosed by solid black lines are LN, whereas outside area is air. The light propagates along $+z$ axis. For $x$-polarization, the crosses represents the $\mathbf{j}_d$ along -$x$ direction, and dots means along +$x$ direction. The sizes of the symbols show the magnitudes. For $y$-polarization case, the instantaneous distributions of displacement current density $\mathbf{j}_d$ are presented by black arrows (length in logarithmic scale).}
\end{figure*}

The idea of metasurfaces may help. The metasurfaces allow to funnel light from the far-field to a deep subwavelength scale, and manipulate light behavior to achieve optical functionalities unprecedented in the natural materials.\cite{yu2011,landy2008} Originally, metasurfaces have been made of plasmonic metals, in which the electromagnetic energy is confined and light-matter interactions are boosted within nanometer scale, leading to giant chirality,\cite{plum2009,qiu2018} enhanced nonlinearities,\cite{kauranen2012,ren2012} and high sensitive sensing.\cite{wu2016} However, the intrinsic Ohmic losses hinder their applications. Recent emerging all-dielectric metasurfaces is a promising alternative to metallic structures, which are believed to alleviate losses, and are more accessible to both electric and magnetic resonances than plasmonic counterparts,\cite{kuznetsov2016}. Until now, various novel optical elements with nanoscale dimensions,\cite{yu2014} such as metalens,\cite{khorasaninejad2016,wangshu2018} waveplates,\cite{lin2014} Huygens surfaces,\cite{decker2015} have been demonstrated.

Here we presented LN as an alternative material to realize metasurfaces in visible and infrared frequency ranges. We fabricated nanograting metasurfaces in the thin crystalline LN film. The nanograting structures are easy to manufacture, which were shown to support superior resonances and new functionalities, such as magnetic mirrors,\cite{liu2017} structural color surface,\cite{zhu2015,gholipour2017}.  As a result of a high refractive index contrast between LN ($n>$2) and ambient media, both electric- and magnetic- dipole moments are well formed inside LN nanostructures. Fano shaped resonances are observed in the transmission spectra as a result of the collective lattice interactions between isolated LN ridges. The spectral positions of the resonances can be engineered by varying the geometrical parameters. Such metasurfaces are proved to exhibit vivid structural colors, which cover the range from pink to purple dependent on different structural dimensions.

\begin{figure} [!t]
\includegraphics[width=60mm]{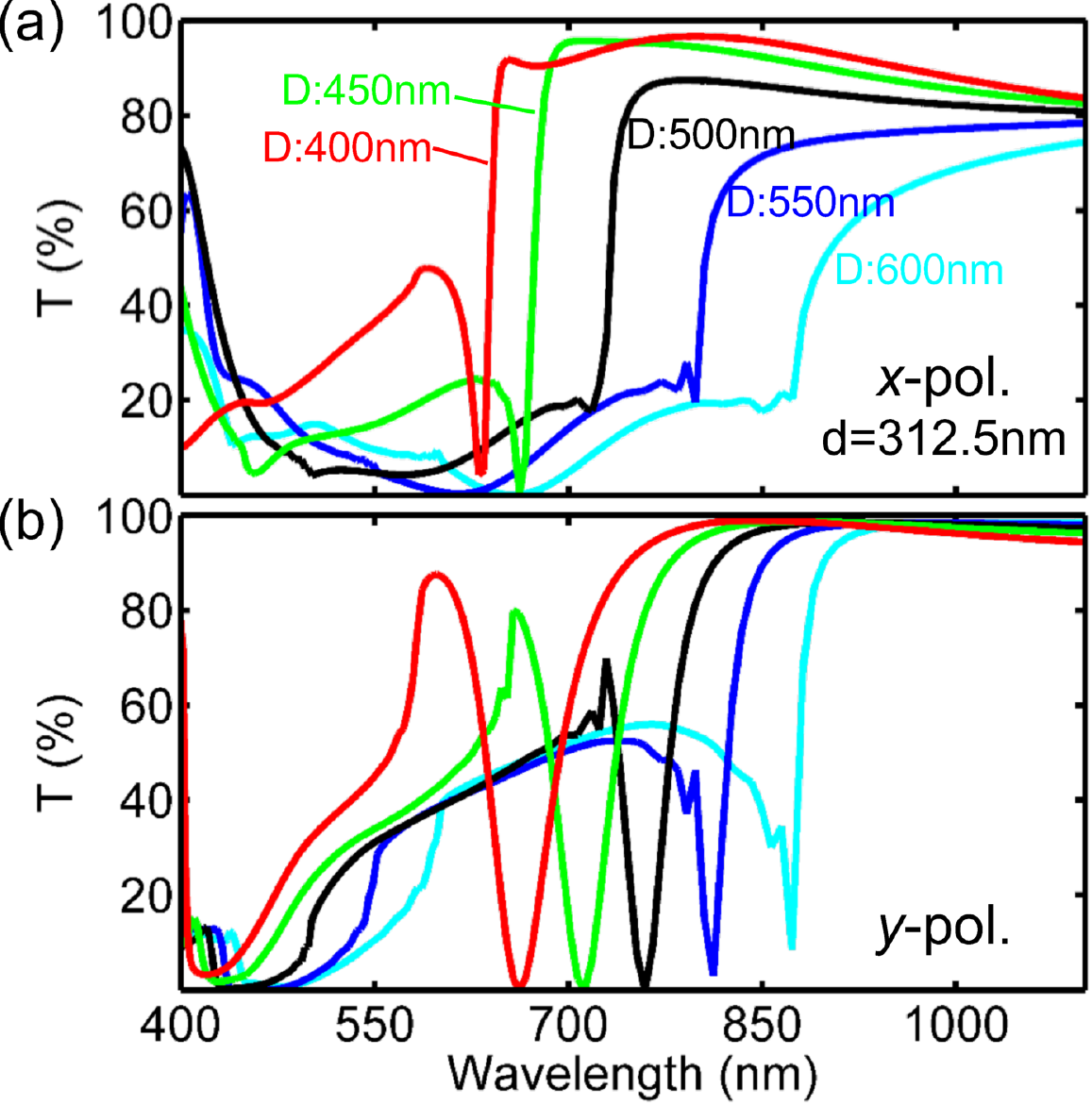} \caption{\label{fig3}
\textbf{Evolution of spectra for metasurfaces with different ridge separations.} \textbf{a}, The transmission spectra for metasurfaces with $d$=312.5~nm unchanged, while $D$ increases from 400 to 600~nm. This is equivalent to increase the separations between the ridges. The real geometric profile of the LN ridge was considered. \textbf{b}, The similar results for $y$-polarization incidence.}
\end{figure}

The representative scanning-electron micrographs (SEM) of nanograting metasurface are shown in Fig.~\ref{fig1}(a). A LNOI wafer (from NANOLN Corporation) with a $x$-cut 220~nm thick LN membrane on top of a silica insulator was adopted. We fabricated the nanograting metasurfaces with different periods ($D$) using focussed ion-beam (FIB, Ga$^{+}$). The duty cycles (percentage of LN ridge width over one period) of each structures remain 62.5\% unchanged. The entire array footprints are $40\times 40~\mathrm{\mu m}^2$. Due to the angle divergence and scattering of the Ga$^{+}$ beams during fabrications, the practical cross-section of the final ridges become round shaped on the top, which deviates ideal rectangle design, as shown by the cross-section image. Figure~\ref{fig1}(b) and (c) show the simulated and experimental transmission of metasurfaces for normal incidence with polarization parallel ($x$-polarization) and perpendicular ($y$-polarization) to the ridges for structures with $D$=400, 500 and 600~nm ($d$=250, 312.5, 375~nm), respectively. Electromagnetic simulations were performed using the finite element method (Comsol Multiphysics). The material parameters of LN were taken from ellipsometric measurements. The light was incident from the far field beneath the sample and output from the air side, which guarantees that the light with wavelength larger than the $D$ would not diffract in the transmitted direction. The round shape of the ridges was considered in the simulation modeling. Both the spectral resonance positions and the lineshape exhibit a high sensitivity to both the geometric parameters of the metasurfaces and the light polarizations. The spectra shift to longer wavelength for larger $D$. For $x$-polarized incidence, the spectra exhibit stepwise asymmetric line profiles, and the transmission becomes smaller for larger $D$. In contrast, $y$-polarized excitation gives distinct narrow valleys with asymmetric Fano shape centered at 650, 760 and 907~nm, respectively. Furthermore, the transmission curves become relative flat and nearly 100\% in a wide spectral range beyond the right slopes of the resonance valleys, which mean the structures come closer to satisfying the impedance matching condition for free space, leading to zero light reflectance. Such high transmission benefit from both the high transparency of LN and the structural designs. And it could be used to realize the broadband antireflective films with few hundred nanometers thick, which are much more compact than the traditional multilayer coating films. The measured spectra (by a commercial microspectrophotometer from IdeaOptics Technologies, Fig.~\ref{fig1}(c)) match the simulation results well in the resonant wavelength positions (as indicated by the dashed lines), except for the lower transmission levels and the reduced quality factors of the resonances. For example, for the $x$-polarized incidence, the step slopes of the measured spectra are more inclined than the simulated ones. And the sharp resonance valley at the blue edge of the spectral step shown in simulation for 400~nm structure becomes broader in experiment. For the $y$-polarization, the resonance valleys are shallow with reduced contrasts. Such differences may result from the LN structure roughness, the Ga$^{+}$ contamination by FIB, all of which would cause extra optical loss and deteriorated optical performance, but were not considered in the simulations.\cite{geiss2014}

The spectral resonances could be explained by the collective interactions between the local responses of each LN ridges. Due to the relatively higher refractive index of LN compared to the surrounding air and silica, each LN ridge with subwavelength width supports Mie resonances, which induce both the electric and the magnetic responses. We numerically evaluate the electric and the magnetic dipole moments excited in a single ridge with $d$=312.5~nm and infinity length along $x$-direction. Using the multipole decomposition approach,\cite{chen2011,evlyukhin2016} the electric and magnetic moments are defined by $\mathbf{p}=\int_{V}{\chi(\mathbf{r})\mathbf{E(r)}dv}$ and $\mathbf{m}=\frac{1}{2}\int_{V}{\mathbf{r}\times\mathbf{J}(\mathbf{r})dv}$, respectively. $\chi(\mathbf{r})$, $\mathbf{E(r)}$, and $\mathbf{J}(\mathbf{r})$ are the electric susceptibility, local electric field, and the displacement currents induced within the ridge at position $\mathbf{r}$ (defined in a Cartesian basis with coordinates’ center coinciding with the ridge's geometric center). The volume integration is performed within the ridge. Considering that the emission of electric dipole is $c$ times larger than their magnetic counterpart with the same magnitude in vacuum,\cite{jackson2012} the \textbf{m} is divided by $c$ for direct comparison of their contributions to the far field radiation. As shown by Fig.~\ref{fig2}(a), both the electric and magnetic dipole moments are excited in an isolate single ridge (shown by inset). The \textbf{p} shows small resonance dips at 466~nm and 677~nm (indicated by the dashed lines) for $x$-polarized light, and a very broad \textbf{p} valley appears around 788~nm for the $y$-polarized excitation. On the other hand, \textbf{m} shows broad resonances centered around 678~nm for the $x$-polarization, while 600~nm and 752~nm for $y$-polarization. However, when the LN ridges are arranged in an ordered array (shown in Fig.~\ref{fig2}(b) inset), the electromagnetic fields of any single ridge may influence the response of neighboring ones. In this way, the individual ridges are coupled together forming collective lattice modes, which facilitates the narrowing and enhancement of the \textbf{p} and \textbf{m} resonances similar to that reported in the plasmonic arrays.\cite{kravets2018} As an example, the results of the periodic array with $D$=500~nm are shown in Fig.~\ref{fig2}(b). For $x$-polarized incidence, both \textbf{p} and \textbf{m} are enhanced dramatically compared to the isolated single ridge response. The unconspicuous resonance dips for the isolated ridge become remarkable distinct and narrow valleys around 488~nm and 656~nm (indicated by the dashed lines connecting Fig.~ref{fig2}(a) and Fig.~ref{fig2}(b)). The stepwise line shape of \textbf{p} spectrum resembles that of the transmission spectrum shown in Fig.~\ref{fig1}. In the mean time, the \textbf{m} resonance is also narrowed and enhanced as a result of the lattice periodicity. For $y$-polarized incidence, the similar enhancement and narrowing effects also happen. In this case, the response is dominated by the magnetic component peaked around 768~nm, where the broad \textbf{p} valley of isolated ridge narrows to presents a obvious dip with minimum near zero. The strong \textbf{m} acts a secondary wave source, leading to a significant backward reflection and a predominant transmission valley in the Fig.~\ref{fig1}.

The resonant responses are further characterized by the hot-spots and displacement currents formed within the grating layer. Figure~\ref{fig2}(c) gives the representative situation of metasurface with $D$=500~nm, and the light is incident along $+z$-direction. For $x$-polarization incidence, the electric field distributions give similar features for three wavelength shown in the figure, in which electric hot-spots mainly localize at the top and bottom of the LN ridge, meanwhile in the air gap between ridges. The magnetic field $H_y$ and induced displacement current $\mathbf{j}_d$ are shown on the right column. The ridges are assumed to have infinite length in the simulations, thus the induced displacement current only has the component parallel to $x$-direction as a result of translational symmetry restriction. At around 737~nm, the $\mathbf{j}_d$ nearly show no circulating features, however, due to the different magnitude of $\mathbf{j}_d$ between the top and bottom of LN ridge, closed line Integral of current is non-zero, hence the effective magnetic response can still be excited following the Ampere's circuital law. At other wavelengths, anti parrallel $\mathbf{j}_d$s are formed, generating magnetic dipole moment. Under $y$-polarization incidence, the magnetic fields show stronger magnitudes than the case of $x$-polarized light at the wavelengths around resonances. This is consistent with the appearance of the circulating displacement currents shown by the arrows in most right column in Fig.~\ref{fig2}(c), which forms strong magnetic dipoles oriented along $x$ at the resonance around 768~nm. In contrast, at other wavelengths, the displacement currents with different circulating directions cancel the formation of \textbf{m}, leading to much smaller \textbf{m} values at those wavelengths.

Such lattice collective resonance is highly sensitive to the separations between the LN ridges. Figure~\ref{fig3} gives the evolution of spectra for metasurfaces with $d$=312.5~nm unchanged, while $D$ increases from 400 to 600~nm, implying the ridge separation enlarges. The spectra suffer dramatical redshift because longer wavelength are needed to match the lattice resonance conditions. In the meantime, the coupling between the neighboring ridges, hence the lattice resonance becomes weaker for larger separation. This cause broadening of the resonances, and reduce their quality factors. Specifically, for $x$-polarization, the transmission level decreases as $D$ increases, and the sharp resonances valley at the blue edge of the spectrum step disappears for larger separations. On the other hand, the resonance valley for $y$-polarization has much lower contrast on the left slope for the case of $D$=600~nm compared to 400~nm, and the minimum of the valley significant deviate from the zero level. 

\begin{figure} [tbp]
\includegraphics[width=80mm]{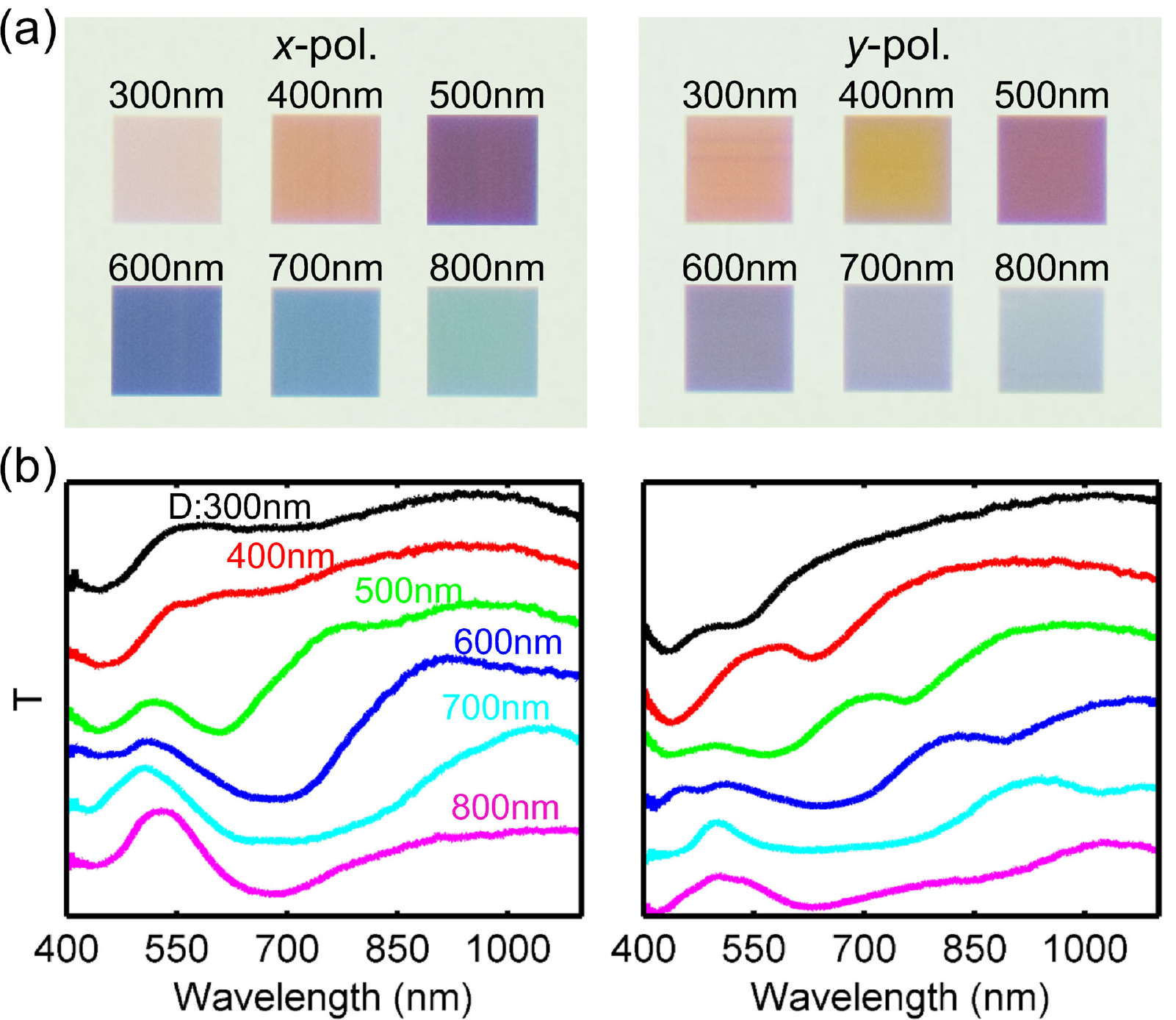} \caption{\label{fig4}
\textbf{The structural colors from the LN metasurfaces.} \textbf{a}, Photographs of LN metasurfaces, illustrating the vibrant colors and the continuous color change with increasing lattice periods from 300 to 800~nm. \textbf{b}, The experimental spectra for the LN metasurfaces under orthogonal polarizations.}
\end{figure}

Such distinct structural resonances of the LN metasurfaces provide an access to a artificial structural colors. As LN has high transparency in the visible spectral range, it is an ideal platform to realize the high efficiency structural colors in transmission mode, which is in remarkable contrast to the reflection colors demonstrated by the plasmonic and Si based metasurfaces.\cite{kristensen2017,lee2018,hu2018} We fabricated a group of metasurfaces with period varied from 300 to 800~nm in 100~nm increment, and illuminated them by white light. The bright field transmission optical microscope images are shown in Fig.~\ref{fig4}(a), whose background gives the color of the unstructured LN film perceived by a CMOS camera. Vivid colors that obviously differ from the background are achieved in the metasurface positions. The photograph Their spectra are shown in Fig.~\ref{fig4}(b). The spectra, and the perceived colors are tunable by both the structural geometries and polarizations. Colors ranging from pink, green, blue to purple could be achieved. Such results could be promisingly applied in the next generation displaying color filters. Furthermore, as a matter of fact, based on various structure designs, the available chromaticity space is unlimited using LN.

In conclusion, we presented the first demonstration of all-dielectric LN metasurfaces. We showed that nanogratings are flexible structures to engineer Mie optical resonances and tune the spectral responses through simple varying the geometric parameters, such as the periodic and separation of the ridges. This also provide a versatile toolset to create structural color palettes over the visible frequency range. Different from the most of the previous metasurfaces fabricated using amorphous materials, our LN metasurfaces were prepared in crystalline film. Thus many distinguishable properties rooted in the lattice symmetry could be maintained in the metasurfaces, such as second order optical nonlinearity, and electro-optical effects, and so forth. This will unambiguously profit the future researches of the switchable and nonlinear dielectric metasurfaces. Furthermore, the LN have wider band gap ($\sim$4~eV) than mostly used metasurface materials, such as silicon ($\sim$1.12~eV), titanium dioxide ($\sim$3~eV), gallium nitride ($\sim$3.4~eV), which benefits nanostructures with low optical loss for wider spectral range. The realization of LN metasurfaces would hold significant promise for the novel functional components for photonics in the future. 


\section{References}
\bibliography{Color_LN_MM}

\begin{thebibliography}{51}%
\makeatletter
\providecommand \@ifxundefined [1]{%
 \@ifx{#1\undefined}
}%
\providecommand \@ifnum [1]{%
 \ifnum #1\expandafter \@firstoftwo
 \else \expandafter \@secondoftwo
 \fi
}%
\providecommand \@ifx [1]{%
 \ifx #1\expandafter \@firstoftwo
 \else \expandafter \@secondoftwo
 \fi
}%
\providecommand \natexlab [1]{#1}%
\providecommand \enquote  [1]{``#1''}%
\providecommand \bibnamefont  [1]{#1}%
\providecommand \bibfnamefont [1]{#1}%
\providecommand \citenamefont [1]{#1}%
\providecommand \href@noop [0]{\@secondoftwo}%
\providecommand \href [0]{\begingroup \@sanitize@url \@href}%
\providecommand \@href[1]{\@@startlink{#1}\@@href}%
\providecommand \@@href[1]{\endgroup#1\@@endlink}%
\providecommand \@sanitize@url [0]{\catcode `\\12\catcode `\$12\catcode
  `\&12\catcode `\#12\catcode `\^12\catcode `\_12\catcode `\%12\relax}%
\providecommand \@@startlink[1]{}%
\providecommand \@@endlink[0]{}%
\providecommand \url  [0]{\begingroup\@sanitize@url \@url }%
\providecommand \@url [1]{\endgroup\@href {#1}{\urlprefix }}%
\providecommand \urlprefix  [0]{URL }%
\providecommand \Eprint [0]{\href }%
\providecommand \doibase [0]{http://dx.doi.org/}%
\providecommand \selectlanguage [0]{\@gobble}%
\providecommand \bibinfo  [0]{\@secondoftwo}%
\providecommand \bibfield  [0]{\@secondoftwo}%
\providecommand \translation [1]{[#1]}%
\providecommand \BibitemOpen [0]{}%
\providecommand \bibitemStop [0]{}%
\providecommand \bibitemNoStop [0]{.\EOS\space}%
\providecommand \EOS [0]{\spacefactor3000\relax}%
\providecommand \BibitemShut  [1]{\csname bibitem#1\endcsname}%
\let\auto@bib@innerbib\@empty
\bibitem [{\citenamefont {Chen}(2012)}]{chen2012}%
  \BibitemOpen
  \bibfield  {author} {\bibinfo {author} {\bibfnamefont {F.}~\bibnamefont
  {Chen}},\ }\href@noop {} {\bibfield  {journal} {\bibinfo  {journal} {Laser \&
  Photon. Rev.}\ }\textbf {\bibinfo {volume} {6}},\ \bibinfo {pages} {622}
  (\bibinfo {year} {2012})}\BibitemShut {NoStop}%
\bibitem [{\citenamefont {Poberaj}\ \emph {et~al.}(2012)\citenamefont
  {Poberaj}, \citenamefont {Hu}, \citenamefont {Sohler},\ and\ \citenamefont
  {Guenter}}]{poberaj2012}%
  \BibitemOpen
  \bibfield  {author} {\bibinfo {author} {\bibfnamefont {G.}~\bibnamefont
  {Poberaj}}, \bibinfo {author} {\bibfnamefont {H.}~\bibnamefont {Hu}},
  \bibinfo {author} {\bibfnamefont {W.}~\bibnamefont {Sohler}}, \ and\ \bibinfo
  {author} {\bibfnamefont {P.}~\bibnamefont {Guenter}},\ }\href@noop {}
  {\bibfield  {journal} {\bibinfo  {journal} {Laser \& Photon. Rev.}\ }\textbf
  {\bibinfo {volume} {6}},\ \bibinfo {pages} {488} (\bibinfo {year}
  {2012})}\BibitemShut {NoStop}%
\bibitem [{\citenamefont {Weis}\ and\ \citenamefont
  {Gaylord}(1985)}]{weis1985}%
  \BibitemOpen
  \bibfield  {author} {\bibinfo {author} {\bibfnamefont {R.}~\bibnamefont
  {Weis}}\ and\ \bibinfo {author} {\bibfnamefont {T.}~\bibnamefont {Gaylord}},\
  }\href@noop {} {\bibfield  {journal} {\bibinfo  {journal} {Appl. Phys. A}\
  }\textbf {\bibinfo {volume} {37}},\ \bibinfo {pages} {191} (\bibinfo {year}
  {1985})}\BibitemShut {NoStop}%
\bibitem [{\citenamefont {Kong}\ \emph {et~al.}(2007)\citenamefont {Kong},
  \citenamefont {Liu}, \citenamefont {Zhao}, \citenamefont {Liu}, \citenamefont
  {Chen},\ and\ \citenamefont {Xu}}]{kong2007}%
  \BibitemOpen
  \bibfield  {author} {\bibinfo {author} {\bibfnamefont {Y.}~\bibnamefont
  {Kong}}, \bibinfo {author} {\bibfnamefont {S.}~\bibnamefont {Liu}}, \bibinfo
  {author} {\bibfnamefont {Y.}~\bibnamefont {Zhao}}, \bibinfo {author}
  {\bibfnamefont {H.}~\bibnamefont {Liu}}, \bibinfo {author} {\bibfnamefont
  {S.}~\bibnamefont {Chen}}, \ and\ \bibinfo {author} {\bibfnamefont
  {J.}~\bibnamefont {Xu}},\ }\href@noop {} {\bibfield  {journal} {\bibinfo
  {journal} {Appl. Phys. Lett.}\ }\textbf {\bibinfo {volume} {91}},\ \bibinfo
  {pages} {081908} (\bibinfo {year} {2007})}\BibitemShut {NoStop}%
\bibitem [{\citenamefont {Kong}, \citenamefont {Liu},\ and\ \citenamefont
  {Xu}(2012)}]{kong2012}%
  \BibitemOpen
  \bibfield  {author} {\bibinfo {author} {\bibfnamefont {Y.}~\bibnamefont
  {Kong}}, \bibinfo {author} {\bibfnamefont {S.}~\bibnamefont {Liu}}, \ and\
  \bibinfo {author} {\bibfnamefont {J.}~\bibnamefont {Xu}},\ }\href@noop {}
  {\bibfield  {journal} {\bibinfo  {journal} {Materials}\ }\textbf {\bibinfo
  {volume} {5}},\ \bibinfo {pages} {1954} (\bibinfo {year} {2012})}\BibitemShut
  {NoStop}%
\bibitem [{\citenamefont {Arizmendi}(2004)}]{arizmendi2004}%
  \BibitemOpen
  \bibfield  {author} {\bibinfo {author} {\bibfnamefont {L.}~\bibnamefont
  {Arizmendi}},\ }\href@noop {} {\bibfield  {journal} {\bibinfo  {journal}
  {Phys. Status Solidi A}\ }\textbf {\bibinfo {volume} {201}},\ \bibinfo
  {pages} {253} (\bibinfo {year} {2004})}\BibitemShut {NoStop}%
\bibitem [{\citenamefont {Wooten}\ \emph {et~al.}(2000)\citenamefont {Wooten},
  \citenamefont {Kissa}, \citenamefont {Yi-Yan}, \citenamefont {Murphy},
  \citenamefont {Lafaw}, \citenamefont {Hallemeier}, \citenamefont {Maack},
  \citenamefont {Attanasio}, \citenamefont {Fritz}, \citenamefont {McBrien}
  \emph {et~al.}}]{wooten2000}%
  \BibitemOpen
  \bibfield  {author} {\bibinfo {author} {\bibfnamefont {E.~L.}\ \bibnamefont
  {Wooten}}, \bibinfo {author} {\bibfnamefont {K.~M.}\ \bibnamefont {Kissa}},
  \bibinfo {author} {\bibfnamefont {A.}~\bibnamefont {Yi-Yan}}, \bibinfo
  {author} {\bibfnamefont {E.~J.}\ \bibnamefont {Murphy}}, \bibinfo {author}
  {\bibfnamefont {D.~A.}\ \bibnamefont {Lafaw}}, \bibinfo {author}
  {\bibfnamefont {P.~F.}\ \bibnamefont {Hallemeier}}, \bibinfo {author}
  {\bibfnamefont {D.}~\bibnamefont {Maack}}, \bibinfo {author} {\bibfnamefont
  {D.~V.}\ \bibnamefont {Attanasio}}, \bibinfo {author} {\bibfnamefont {D.~J.}\
  \bibnamefont {Fritz}}, \bibinfo {author} {\bibfnamefont {G.~J.}\ \bibnamefont
  {McBrien}},  \emph {et~al.},\ }\href@noop {} {\bibfield  {journal} {\bibinfo
  {journal} {IEEE J. Sel. Top. Quant.}\ }\textbf {\bibinfo {volume} {6}},\
  \bibinfo {pages} {69} (\bibinfo {year} {2000})}\BibitemShut {NoStop}%
\bibitem [{\citenamefont {Boes}\ \emph {et~al.}(2018)\citenamefont {Boes},
  \citenamefont {Corcoran}, \citenamefont {Chang}, \citenamefont {Bowers},\
  and\ \citenamefont {Mitchell}}]{boes2018}%
  \BibitemOpen
  \bibfield  {author} {\bibinfo {author} {\bibfnamefont {A.}~\bibnamefont
  {Boes}}, \bibinfo {author} {\bibfnamefont {B.}~\bibnamefont {Corcoran}},
  \bibinfo {author} {\bibfnamefont {L.}~\bibnamefont {Chang}}, \bibinfo
  {author} {\bibfnamefont {J.}~\bibnamefont {Bowers}}, \ and\ \bibinfo {author}
  {\bibfnamefont {A.}~\bibnamefont {Mitchell}},\ }\href@noop {} {\bibfield
  {journal} {\bibinfo  {journal} {Laser \& Photon. Rev.}\ }\textbf {\bibinfo
  {volume} {12}},\ \bibinfo {pages} {1700256} (\bibinfo {year}
  {2018})}\BibitemShut {NoStop}%
\bibitem [{\citenamefont {Kip}(1998)}]{kip1998}%
  \BibitemOpen
  \bibfield  {author} {\bibinfo {author} {\bibfnamefont {D.}~\bibnamefont
  {Kip}},\ }\href@noop {} {\bibfield  {journal} {\bibinfo  {journal} {Appl.
  Phys. B}\ }\textbf {\bibinfo {volume} {67}},\ \bibinfo {pages} {131}
  (\bibinfo {year} {1998})}\BibitemShut {NoStop}%
\bibitem [{\citenamefont {Lin}\ \emph {et~al.}(2015)\citenamefont {Lin},
  \citenamefont {Xu}, \citenamefont {Fang}, \citenamefont {Wang}, \citenamefont
  {Song}, \citenamefont {Wang}, \citenamefont {Qiao}, \citenamefont {Fang},\
  and\ \citenamefont {Cheng}}]{lin2015}%
  \BibitemOpen
  \bibfield  {author} {\bibinfo {author} {\bibfnamefont {J.}~\bibnamefont
  {Lin}}, \bibinfo {author} {\bibfnamefont {Y.}~\bibnamefont {Xu}}, \bibinfo
  {author} {\bibfnamefont {Z.}~\bibnamefont {Fang}}, \bibinfo {author}
  {\bibfnamefont {M.}~\bibnamefont {Wang}}, \bibinfo {author} {\bibfnamefont
  {J.}~\bibnamefont {Song}}, \bibinfo {author} {\bibfnamefont {N.}~\bibnamefont
  {Wang}}, \bibinfo {author} {\bibfnamefont {L.}~\bibnamefont {Qiao}}, \bibinfo
  {author} {\bibfnamefont {W.}~\bibnamefont {Fang}}, \ and\ \bibinfo {author}
  {\bibfnamefont {Y.}~\bibnamefont {Cheng}},\ }\href@noop {} {\bibfield
  {journal} {\bibinfo  {journal} {Sci. Rep.}\ }\textbf {\bibinfo {volume}
  {5}},\ \bibinfo {pages} {8072} (\bibinfo {year} {2015})}\BibitemShut
  {NoStop}%
\bibitem [{\citenamefont {Si}\ \emph {et~al.}(2011)\citenamefont {Si},
  \citenamefont {Danner}, \citenamefont {Teo}, \citenamefont {Teo},
  \citenamefont {Teng},\ and\ \citenamefont {Bettiol}}]{si2011}%
  \BibitemOpen
  \bibfield  {author} {\bibinfo {author} {\bibfnamefont {G.}~\bibnamefont
  {Si}}, \bibinfo {author} {\bibfnamefont {A.~J.}\ \bibnamefont {Danner}},
  \bibinfo {author} {\bibfnamefont {S.~L.}\ \bibnamefont {Teo}}, \bibinfo
  {author} {\bibfnamefont {E.~J.}\ \bibnamefont {Teo}}, \bibinfo {author}
  {\bibfnamefont {J.}~\bibnamefont {Teng}}, \ and\ \bibinfo {author}
  {\bibfnamefont {A.~A.}\ \bibnamefont {Bettiol}},\ }\href@noop {} {\bibfield
  {journal} {\bibinfo  {journal} {Vac. Sci. Technol. B}\ }\textbf {\bibinfo
  {volume} {29}},\ \bibinfo {pages} {021205} (\bibinfo {year}
  {2011})}\BibitemShut {NoStop}%
\bibitem [{\citenamefont {Chen}(2009)}]{chen2009}%
  \BibitemOpen
  \bibfield  {author} {\bibinfo {author} {\bibfnamefont {F.}~\bibnamefont
  {Chen}},\ }\href@noop {} {\bibfield  {journal} {\bibinfo  {journal} {J. Appl.
  Phys.}\ }\textbf {\bibinfo {volume} {106}},\ \bibinfo {pages} {11} (\bibinfo
  {year} {2009})}\BibitemShut {NoStop}%
\bibitem [{\citenamefont {Courjal}\ \emph {et~al.}(2011)\citenamefont
  {Courjal}, \citenamefont {Guichardaz}, \citenamefont {Ulliac}, \citenamefont
  {Rauch}, \citenamefont {Sadani}, \citenamefont {Lu},\ and\ \citenamefont
  {Bernal}}]{courjal2011}%
  \BibitemOpen
  \bibfield  {author} {\bibinfo {author} {\bibfnamefont {N.}~\bibnamefont
  {Courjal}}, \bibinfo {author} {\bibfnamefont {B.}~\bibnamefont {Guichardaz}},
  \bibinfo {author} {\bibfnamefont {G.}~\bibnamefont {Ulliac}}, \bibinfo
  {author} {\bibfnamefont {J.-Y.}\ \bibnamefont {Rauch}}, \bibinfo {author}
  {\bibfnamefont {B.}~\bibnamefont {Sadani}}, \bibinfo {author} {\bibfnamefont
  {H.-H.}\ \bibnamefont {Lu}}, \ and\ \bibinfo {author} {\bibfnamefont {M.-P.}\
  \bibnamefont {Bernal}},\ }\href@noop {} {\bibfield  {journal} {\bibinfo
  {journal} {J. Phys. D}\ }\textbf {\bibinfo {volume} {44}},\ \bibinfo {pages}
  {305101} (\bibinfo {year} {2011})}\BibitemShut {NoStop}%
\bibitem [{\citenamefont {Bazzan}\ and\ \citenamefont
  {Sada}(2015)}]{bazzan2015}%
  \BibitemOpen
  \bibfield  {author} {\bibinfo {author} {\bibfnamefont {M.}~\bibnamefont
  {Bazzan}}\ and\ \bibinfo {author} {\bibfnamefont {C.}~\bibnamefont {Sada}},\
  }\href@noop {} {\bibfield  {journal} {\bibinfo  {journal} {Appl. Phys. Rev.}\
  }\textbf {\bibinfo {volume} {2}},\ \bibinfo {pages} {040603} (\bibinfo {year}
  {2015})}\BibitemShut {NoStop}%
\bibitem [{\citenamefont {Bernal}\ \emph {et~al.}(2006)\citenamefont {Bernal},
  \citenamefont {Courjal}, \citenamefont {Amet}, \citenamefont {Roussey},\ and\
  \citenamefont {Hou}}]{bernal2006}%
  \BibitemOpen
  \bibfield  {author} {\bibinfo {author} {\bibfnamefont {M.-P.}\ \bibnamefont
  {Bernal}}, \bibinfo {author} {\bibfnamefont {N.}~\bibnamefont {Courjal}},
  \bibinfo {author} {\bibfnamefont {J.}~\bibnamefont {Amet}}, \bibinfo {author}
  {\bibfnamefont {M.}~\bibnamefont {Roussey}}, \ and\ \bibinfo {author}
  {\bibfnamefont {C.}~\bibnamefont {Hou}},\ }\href@noop {} {\bibfield
  {journal} {\bibinfo  {journal} {Opt. Commun.}\ }\textbf {\bibinfo {volume}
  {265}},\ \bibinfo {pages} {180} (\bibinfo {year} {2006})}\BibitemShut
  {NoStop}%
\bibitem [{\citenamefont {Diziain}\ \emph {et~al.}(2008)\citenamefont
  {Diziain}, \citenamefont {Amet}, \citenamefont {Baida},\ and\ \citenamefont
  {Bernal}}]{diziain2008}%
  \BibitemOpen
  \bibfield  {author} {\bibinfo {author} {\bibfnamefont {S.}~\bibnamefont
  {Diziain}}, \bibinfo {author} {\bibfnamefont {J.}~\bibnamefont {Amet}},
  \bibinfo {author} {\bibfnamefont {F.~I.}\ \bibnamefont {Baida}}, \ and\
  \bibinfo {author} {\bibfnamefont {M.-P.}\ \bibnamefont {Bernal}},\
  }\href@noop {} {\bibfield  {journal} {\bibinfo  {journal} {Appl. Phys.
  Lett.}\ }\textbf {\bibinfo {volume} {93}},\ \bibinfo {pages} {261103}
  (\bibinfo {year} {2008})}\BibitemShut {NoStop}%
\bibitem [{\citenamefont {Sulser}\ \emph {et~al.}(2009)\citenamefont {Sulser},
  \citenamefont {Poberaj}, \citenamefont {Koechlin},\ and\ \citenamefont
  {G{\"u}nter}}]{sulser2009}%
  \BibitemOpen
  \bibfield  {author} {\bibinfo {author} {\bibfnamefont {F.}~\bibnamefont
  {Sulser}}, \bibinfo {author} {\bibfnamefont {G.}~\bibnamefont {Poberaj}},
  \bibinfo {author} {\bibfnamefont {M.}~\bibnamefont {Koechlin}}, \ and\
  \bibinfo {author} {\bibfnamefont {P.}~\bibnamefont {G{\"u}nter}},\
  }\href@noop {} {\bibfield  {journal} {\bibinfo  {journal} {Opt. Express}\
  }\textbf {\bibinfo {volume} {17}},\ \bibinfo {pages} {20291} (\bibinfo {year}
  {2009})}\BibitemShut {NoStop}%
\bibitem [{\citenamefont {Geiss}\ \emph {et~al.}(2010)\citenamefont {Geiss},
  \citenamefont {Diziain}, \citenamefont {Iliew}, \citenamefont {Etrich},
  \citenamefont {Hartung}, \citenamefont {Janunts}, \citenamefont {Schrempel},
  \citenamefont {Lederer}, \citenamefont {Pertsch},\ and\ \citenamefont
  {Kley}}]{geiss2010}%
  \BibitemOpen
  \bibfield  {author} {\bibinfo {author} {\bibfnamefont {R.}~\bibnamefont
  {Geiss}}, \bibinfo {author} {\bibfnamefont {S.}~\bibnamefont {Diziain}},
  \bibinfo {author} {\bibfnamefont {R.}~\bibnamefont {Iliew}}, \bibinfo
  {author} {\bibfnamefont {C.}~\bibnamefont {Etrich}}, \bibinfo {author}
  {\bibfnamefont {H.}~\bibnamefont {Hartung}}, \bibinfo {author} {\bibfnamefont
  {N.}~\bibnamefont {Janunts}}, \bibinfo {author} {\bibfnamefont
  {F.}~\bibnamefont {Schrempel}}, \bibinfo {author} {\bibfnamefont
  {F.}~\bibnamefont {Lederer}}, \bibinfo {author} {\bibfnamefont
  {T.}~\bibnamefont {Pertsch}}, \ and\ \bibinfo {author} {\bibfnamefont
  {E.-B.}\ \bibnamefont {Kley}},\ }\href@noop {} {\bibfield  {journal}
  {\bibinfo  {journal} {Appl. Phys. Lett.}\ }\textbf {\bibinfo {volume} {97}},\
  \bibinfo {pages} {131109} (\bibinfo {year} {2010})}\BibitemShut {NoStop}%
\bibitem [{\citenamefont {Cai}\ \emph {et~al.}(2014)\citenamefont {Cai},
  \citenamefont {Han}, \citenamefont {Zhang}, \citenamefont {Hu},\ and\
  \citenamefont {Wang}}]{cai2014}%
  \BibitemOpen
  \bibfield  {author} {\bibinfo {author} {\bibfnamefont {L.}~\bibnamefont
  {Cai}}, \bibinfo {author} {\bibfnamefont {H.}~\bibnamefont {Han}}, \bibinfo
  {author} {\bibfnamefont {S.}~\bibnamefont {Zhang}}, \bibinfo {author}
  {\bibfnamefont {H.}~\bibnamefont {Hu}}, \ and\ \bibinfo {author}
  {\bibfnamefont {K.}~\bibnamefont {Wang}},\ }\href@noop {} {\bibfield
  {journal} {\bibinfo  {journal} {Opt. Lett.}\ }\textbf {\bibinfo {volume}
  {39}},\ \bibinfo {pages} {2094} (\bibinfo {year} {2014})}\BibitemShut
  {NoStop}%
\bibitem [{\citenamefont {Diziain}\ \emph {et~al.}(2015)\citenamefont
  {Diziain}, \citenamefont {Geiss}, \citenamefont {Steinert}, \citenamefont
  {Schmidt}, \citenamefont {Chang}, \citenamefont {Fasold}, \citenamefont
  {F{\"u}{\ss}el}, \citenamefont {Chen},\ and\ \citenamefont
  {Pertsch}}]{diziain2015}%
  \BibitemOpen
  \bibfield  {author} {\bibinfo {author} {\bibfnamefont {S.}~\bibnamefont
  {Diziain}}, \bibinfo {author} {\bibfnamefont {R.}~\bibnamefont {Geiss}},
  \bibinfo {author} {\bibfnamefont {M.}~\bibnamefont {Steinert}}, \bibinfo
  {author} {\bibfnamefont {C.}~\bibnamefont {Schmidt}}, \bibinfo {author}
  {\bibfnamefont {W.-K.}\ \bibnamefont {Chang}}, \bibinfo {author}
  {\bibfnamefont {S.}~\bibnamefont {Fasold}}, \bibinfo {author} {\bibfnamefont
  {D.}~\bibnamefont {F{\"u}{\ss}el}}, \bibinfo {author} {\bibfnamefont {Y.-H.}\
  \bibnamefont {Chen}}, \ and\ \bibinfo {author} {\bibfnamefont
  {T.}~\bibnamefont {Pertsch}},\ }\href@noop {} {\bibfield  {journal} {\bibinfo
   {journal} {Opt. Mater. Express}\ }\textbf {\bibinfo {volume} {5}},\ \bibinfo
  {pages} {2081} (\bibinfo {year} {2015})}\BibitemShut {NoStop}%
\bibitem [{\citenamefont {Wang}\ \emph {et~al.}(2015)\citenamefont {Wang},
  \citenamefont {Bo}, \citenamefont {Wan}, \citenamefont {Li}, \citenamefont
  {Gao}, \citenamefont {Li}, \citenamefont {Zhang},\ and\ \citenamefont
  {Xu}}]{wang2015}%
  \BibitemOpen
  \bibfield  {author} {\bibinfo {author} {\bibfnamefont {J.}~\bibnamefont
  {Wang}}, \bibinfo {author} {\bibfnamefont {F.}~\bibnamefont {Bo}}, \bibinfo
  {author} {\bibfnamefont {S.}~\bibnamefont {Wan}}, \bibinfo {author}
  {\bibfnamefont {W.}~\bibnamefont {Li}}, \bibinfo {author} {\bibfnamefont
  {F.}~\bibnamefont {Gao}}, \bibinfo {author} {\bibfnamefont {J.}~\bibnamefont
  {Li}}, \bibinfo {author} {\bibfnamefont {G.}~\bibnamefont {Zhang}}, \ and\
  \bibinfo {author} {\bibfnamefont {J.}~\bibnamefont {Xu}},\ }\href@noop {}
  {\bibfield  {journal} {\bibinfo  {journal} {Opt. Express}\ }\textbf {\bibinfo
  {volume} {23}},\ \bibinfo {pages} {23072} (\bibinfo {year}
  {2015})}\BibitemShut {NoStop}%
\bibitem [{\citenamefont {Wang}\ \emph
  {et~al.}(2018{\natexlab{a}})\citenamefont {Wang}, \citenamefont {Wang},
  \citenamefont {Wang}, \citenamefont {Bo}, \citenamefont {Zhang},
  \citenamefont {Gong}, \citenamefont {Lon{\v{c}}ar},\ and\ \citenamefont
  {Xiao}}]{wang2018}%
  \BibitemOpen
  \bibfield  {author} {\bibinfo {author} {\bibfnamefont {L.}~\bibnamefont
  {Wang}}, \bibinfo {author} {\bibfnamefont {C.}~\bibnamefont {Wang}}, \bibinfo
  {author} {\bibfnamefont {J.}~\bibnamefont {Wang}}, \bibinfo {author}
  {\bibfnamefont {F.}~\bibnamefont {Bo}}, \bibinfo {author} {\bibfnamefont
  {M.}~\bibnamefont {Zhang}}, \bibinfo {author} {\bibfnamefont
  {Q.}~\bibnamefont {Gong}}, \bibinfo {author} {\bibfnamefont {M.}~\bibnamefont
  {Lon{\v{c}}ar}}, \ and\ \bibinfo {author} {\bibfnamefont {Y.-F.}\
  \bibnamefont {Xiao}},\ }\href@noop {} {\bibfield  {journal} {\bibinfo
  {journal} {Opt. Lett.}\ }\textbf {\bibinfo {volume} {43}},\ \bibinfo {pages}
  {2917} (\bibinfo {year} {2018}{\natexlab{a}})}\BibitemShut {NoStop}%
\bibitem [{\citenamefont {Diziain}\ \emph {et~al.}(2013)\citenamefont
  {Diziain}, \citenamefont {Geiss}, \citenamefont {Zilk}, \citenamefont
  {Schrempel}, \citenamefont {Kley}, \citenamefont {T{\"u}nnermann},\ and\
  \citenamefont {Pertsch}}]{diziain2013}%
  \BibitemOpen
  \bibfield  {author} {\bibinfo {author} {\bibfnamefont {S.}~\bibnamefont
  {Diziain}}, \bibinfo {author} {\bibfnamefont {R.}~\bibnamefont {Geiss}},
  \bibinfo {author} {\bibfnamefont {M.}~\bibnamefont {Zilk}}, \bibinfo {author}
  {\bibfnamefont {F.}~\bibnamefont {Schrempel}}, \bibinfo {author}
  {\bibfnamefont {E.-B.}\ \bibnamefont {Kley}}, \bibinfo {author}
  {\bibfnamefont {A.}~\bibnamefont {T{\"u}nnermann}}, \ and\ \bibinfo {author}
  {\bibfnamefont {T.}~\bibnamefont {Pertsch}},\ }\href@noop {} {\bibfield
  {journal} {\bibinfo  {journal} {Appl. Phys. Lett.}\ }\textbf {\bibinfo
  {volume} {103}},\ \bibinfo {pages} {051117} (\bibinfo {year}
  {2013})}\BibitemShut {NoStop}%
\bibitem [{\citenamefont {Geiss}\ \emph {et~al.}(2015)\citenamefont {Geiss},
  \citenamefont {Saravi}, \citenamefont {Sergeyev}, \citenamefont {Diziain},
  \citenamefont {Setzpfandt}, \citenamefont {Schrempel}, \citenamefont
  {Grange}, \citenamefont {Kley}, \citenamefont {T{\"u}nnermann},\ and\
  \citenamefont {Pertsch}}]{geiss2015}%
  \BibitemOpen
  \bibfield  {author} {\bibinfo {author} {\bibfnamefont {R.}~\bibnamefont
  {Geiss}}, \bibinfo {author} {\bibfnamefont {S.}~\bibnamefont {Saravi}},
  \bibinfo {author} {\bibfnamefont {A.}~\bibnamefont {Sergeyev}}, \bibinfo
  {author} {\bibfnamefont {S.}~\bibnamefont {Diziain}}, \bibinfo {author}
  {\bibfnamefont {F.}~\bibnamefont {Setzpfandt}}, \bibinfo {author}
  {\bibfnamefont {F.}~\bibnamefont {Schrempel}}, \bibinfo {author}
  {\bibfnamefont {R.}~\bibnamefont {Grange}}, \bibinfo {author} {\bibfnamefont
  {E.-B.}\ \bibnamefont {Kley}}, \bibinfo {author} {\bibfnamefont
  {A.}~\bibnamefont {T{\"u}nnermann}}, \ and\ \bibinfo {author} {\bibfnamefont
  {T.}~\bibnamefont {Pertsch}},\ }\href@noop {} {\bibfield  {journal} {\bibinfo
   {journal} {Opt. Lett.}\ }\textbf {\bibinfo {volume} {40}},\ \bibinfo {pages}
  {2715} (\bibinfo {year} {2015})}\BibitemShut {NoStop}%
\bibitem [{\citenamefont {Wang}\ \emph {et~al.}(2017)\citenamefont {Wang},
  \citenamefont {Xiong}, \citenamefont {Andrade}, \citenamefont {Venkataraman},
  \citenamefont {Ren}, \citenamefont {Guo},\ and\ \citenamefont
  {Lon{\v{c}}ar}}]{wang2017}%
  \BibitemOpen
  \bibfield  {author} {\bibinfo {author} {\bibfnamefont {C.}~\bibnamefont
  {Wang}}, \bibinfo {author} {\bibfnamefont {X.}~\bibnamefont {Xiong}},
  \bibinfo {author} {\bibfnamefont {N.}~\bibnamefont {Andrade}}, \bibinfo
  {author} {\bibfnamefont {V.}~\bibnamefont {Venkataraman}}, \bibinfo {author}
  {\bibfnamefont {X.-F.}\ \bibnamefont {Ren}}, \bibinfo {author} {\bibfnamefont
  {G.-C.}\ \bibnamefont {Guo}}, \ and\ \bibinfo {author} {\bibfnamefont
  {M.}~\bibnamefont {Lon{\v{c}}ar}},\ }\href@noop {} {\bibfield  {journal}
  {\bibinfo  {journal} {Opt. Express}\ }\textbf {\bibinfo {volume} {25}},\
  \bibinfo {pages} {6963} (\bibinfo {year} {2017})}\BibitemShut {NoStop}%
\bibitem [{\citenamefont {Lu}\ \emph {et~al.}(2014)\citenamefont {Lu},
  \citenamefont {Qiu}, \citenamefont {Guyot}, \citenamefont {Ulliac},
  \citenamefont {Merolla}, \citenamefont {Baida},\ and\ \citenamefont
  {Bernal}}]{lu2014}%
  \BibitemOpen
  \bibfield  {author} {\bibinfo {author} {\bibfnamefont {H.}~\bibnamefont
  {Lu}}, \bibinfo {author} {\bibfnamefont {W.}~\bibnamefont {Qiu}}, \bibinfo
  {author} {\bibfnamefont {C.}~\bibnamefont {Guyot}}, \bibinfo {author}
  {\bibfnamefont {G.}~\bibnamefont {Ulliac}}, \bibinfo {author} {\bibfnamefont
  {J.-M.}\ \bibnamefont {Merolla}}, \bibinfo {author} {\bibfnamefont
  {F.}~\bibnamefont {Baida}}, \ and\ \bibinfo {author} {\bibfnamefont {M.-P.}\
  \bibnamefont {Bernal}},\ }\href@noop {} {\bibfield  {journal} {\bibinfo
  {journal} {IEEE Photon. Technol. Lett}\ }\textbf {\bibinfo {volume} {26}},\
  \bibinfo {pages} {1332} (\bibinfo {year} {2014})}\BibitemShut {NoStop}%
\bibitem [{\citenamefont {Wang}\ \emph
  {et~al.}(2018{\natexlab{b}})\citenamefont {Wang}, \citenamefont {Zhang},
  \citenamefont {Chen}, \citenamefont {Bertrand}, \citenamefont {Shams-Ansari},
  \citenamefont {Chandrasekhar}, \citenamefont {Winzer},\ and\ \citenamefont
  {Lon{\v{c}}ar}}]{wangcheng2018}%
  \BibitemOpen
  \bibfield  {author} {\bibinfo {author} {\bibfnamefont {C.}~\bibnamefont
  {Wang}}, \bibinfo {author} {\bibfnamefont {M.}~\bibnamefont {Zhang}},
  \bibinfo {author} {\bibfnamefont {X.}~\bibnamefont {Chen}}, \bibinfo {author}
  {\bibfnamefont {M.}~\bibnamefont {Bertrand}}, \bibinfo {author}
  {\bibfnamefont {A.}~\bibnamefont {Shams-Ansari}}, \bibinfo {author}
  {\bibfnamefont {S.}~\bibnamefont {Chandrasekhar}}, \bibinfo {author}
  {\bibfnamefont {P.}~\bibnamefont {Winzer}}, \ and\ \bibinfo {author}
  {\bibfnamefont {M.}~\bibnamefont {Lon{\v{c}}ar}},\ }\href@noop {} {\bibfield
  {journal} {\bibinfo  {journal} {Nature}\ ,\ \bibinfo {pages} {1}} (\bibinfo
  {year} {2018}{\natexlab{b}})}\BibitemShut {NoStop}%
\bibitem [{\citenamefont {Yu}\ \emph {et~al.}(2011)\citenamefont {Yu},
  \citenamefont {Genevet}, \citenamefont {Kats}, \citenamefont {Aieta},
  \citenamefont {Tetienne}, \citenamefont {Capasso},\ and\ \citenamefont
  {Gaburro}}]{yu2011}%
  \BibitemOpen
  \bibfield  {author} {\bibinfo {author} {\bibfnamefont {N.}~\bibnamefont
  {Yu}}, \bibinfo {author} {\bibfnamefont {P.}~\bibnamefont {Genevet}},
  \bibinfo {author} {\bibfnamefont {M.~A.}\ \bibnamefont {Kats}}, \bibinfo
  {author} {\bibfnamefont {F.}~\bibnamefont {Aieta}}, \bibinfo {author}
  {\bibfnamefont {J.-P.}\ \bibnamefont {Tetienne}}, \bibinfo {author}
  {\bibfnamefont {F.}~\bibnamefont {Capasso}}, \ and\ \bibinfo {author}
  {\bibfnamefont {Z.}~\bibnamefont {Gaburro}},\ }\href@noop {} {\bibfield
  {journal} {\bibinfo  {journal} {science}\ ,\ \bibinfo {pages} {1210713}}
  (\bibinfo {year} {2011})}\BibitemShut {NoStop}%
\bibitem [{\citenamefont {Landy}\ \emph {et~al.}(2008)\citenamefont {Landy},
  \citenamefont {Sajuyigbe}, \citenamefont {Mock}, \citenamefont {Smith},\ and\
  \citenamefont {Padilla}}]{landy2008}%
  \BibitemOpen
  \bibfield  {author} {\bibinfo {author} {\bibfnamefont {N.~I.}\ \bibnamefont
  {Landy}}, \bibinfo {author} {\bibfnamefont {S.}~\bibnamefont {Sajuyigbe}},
  \bibinfo {author} {\bibfnamefont {J.~J.}\ \bibnamefont {Mock}}, \bibinfo
  {author} {\bibfnamefont {D.~R.}\ \bibnamefont {Smith}}, \ and\ \bibinfo
  {author} {\bibfnamefont {W.~J.}\ \bibnamefont {Padilla}},\ }\href@noop {}
  {\bibfield  {journal} {\bibinfo  {journal} {Phys. Rev. Lett.}\ }\textbf
  {\bibinfo {volume} {100}},\ \bibinfo {pages} {207402} (\bibinfo {year}
  {2008})}\BibitemShut {NoStop}%
\bibitem [{\citenamefont {Plum}\ \emph {et~al.}(2009)\citenamefont {Plum},
  \citenamefont {Liu}, \citenamefont {Fedotov}, \citenamefont {Chen},
  \citenamefont {Tsai},\ and\ \citenamefont {Zheludev}}]{plum2009}%
  \BibitemOpen
  \bibfield  {author} {\bibinfo {author} {\bibfnamefont {E.}~\bibnamefont
  {Plum}}, \bibinfo {author} {\bibfnamefont {X.-X.}\ \bibnamefont {Liu}},
  \bibinfo {author} {\bibfnamefont {V.}~\bibnamefont {Fedotov}}, \bibinfo
  {author} {\bibfnamefont {Y.}~\bibnamefont {Chen}}, \bibinfo {author}
  {\bibfnamefont {D.}~\bibnamefont {Tsai}}, \ and\ \bibinfo {author}
  {\bibfnamefont {N.}~\bibnamefont {Zheludev}},\ }\href@noop {} {\bibfield
  {journal} {\bibinfo  {journal} {Phys. Rev. Lett.}\ }\textbf {\bibinfo
  {volume} {102}},\ \bibinfo {pages} {113902} (\bibinfo {year}
  {2009})}\BibitemShut {NoStop}%
\bibitem [{\citenamefont {Qiu}\ \emph {et~al.}(2018)\citenamefont {Qiu},
  \citenamefont {Zhang}, \citenamefont {Tang}, \citenamefont {Jin},
  \citenamefont {Qiu},\ and\ \citenamefont {Lei}}]{qiu2018}%
  \BibitemOpen
  \bibfield  {author} {\bibinfo {author} {\bibfnamefont {M.}~\bibnamefont
  {Qiu}}, \bibinfo {author} {\bibfnamefont {L.}~\bibnamefont {Zhang}}, \bibinfo
  {author} {\bibfnamefont {Z.}~\bibnamefont {Tang}}, \bibinfo {author}
  {\bibfnamefont {W.}~\bibnamefont {Jin}}, \bibinfo {author} {\bibfnamefont
  {C.-W.}\ \bibnamefont {Qiu}}, \ and\ \bibinfo {author} {\bibfnamefont
  {D.~Y.}\ \bibnamefont {Lei}},\ }\href@noop {} {\bibfield  {journal} {\bibinfo
   {journal} {Adv. Func. Mater.}\ ,\ \bibinfo {pages} {1803147}} (\bibinfo
  {year} {2018})}\BibitemShut {NoStop}%
\bibitem [{\citenamefont {Kauranen}\ and\ \citenamefont
  {Zayats}(2012)}]{kauranen2012}%
  \BibitemOpen
  \bibfield  {author} {\bibinfo {author} {\bibfnamefont {M.}~\bibnamefont
  {Kauranen}}\ and\ \bibinfo {author} {\bibfnamefont {A.~V.}\ \bibnamefont
  {Zayats}},\ }\href@noop {} {\bibfield  {journal} {\bibinfo  {journal} {Nature
  Photon.}\ }\textbf {\bibinfo {volume} {6}},\ \bibinfo {pages} {737} (\bibinfo
  {year} {2012})}\BibitemShut {NoStop}%
\bibitem [{\citenamefont {Ren}\ \emph {et~al.}(2012)\citenamefont {Ren},
  \citenamefont {Plum}, \citenamefont {Xu},\ and\ \citenamefont
  {Zheludev}}]{ren2012}%
  \BibitemOpen
  \bibfield  {author} {\bibinfo {author} {\bibfnamefont {M.}~\bibnamefont
  {Ren}}, \bibinfo {author} {\bibfnamefont {E.}~\bibnamefont {Plum}}, \bibinfo
  {author} {\bibfnamefont {J.}~\bibnamefont {Xu}}, \ and\ \bibinfo {author}
  {\bibfnamefont {N.~I.}\ \bibnamefont {Zheludev}},\ }\href@noop {} {\bibfield
  {journal} {\bibinfo  {journal} {Nature Commun.}\ }\textbf {\bibinfo {volume}
  {3}},\ \bibinfo {pages} {833} (\bibinfo {year} {2012})}\BibitemShut {NoStop}%
\bibitem [{\citenamefont {Wu}\ \emph {et~al.}(2016)\citenamefont {Wu},
  \citenamefont {Ren}, \citenamefont {Pi}, \citenamefont {Cai},\ and\
  \citenamefont {Xu}}]{wu2016}%
  \BibitemOpen
  \bibfield  {author} {\bibinfo {author} {\bibfnamefont {W.}~\bibnamefont
  {Wu}}, \bibinfo {author} {\bibfnamefont {M.}~\bibnamefont {Ren}}, \bibinfo
  {author} {\bibfnamefont {B.}~\bibnamefont {Pi}}, \bibinfo {author}
  {\bibfnamefont {W.}~\bibnamefont {Cai}}, \ and\ \bibinfo {author}
  {\bibfnamefont {J.}~\bibnamefont {Xu}},\ }\href@noop {} {\bibfield  {journal}
  {\bibinfo  {journal} {Appl. Phys. Lett.}\ }\textbf {\bibinfo {volume}
  {108}},\ \bibinfo {pages} {073106} (\bibinfo {year} {2016})}\BibitemShut
  {NoStop}%
\bibitem [{\citenamefont {Kuznetsov}\ \emph {et~al.}(2016)\citenamefont
  {Kuznetsov}, \citenamefont {Miroshnichenko}, \citenamefont {Brongersma},
  \citenamefont {Kivshar},\ and\ \citenamefont {Luk¡¯yanchuk}}]{kuznetsov2016}%
  \BibitemOpen
  \bibfield  {author} {\bibinfo {author} {\bibfnamefont {A.~I.}\ \bibnamefont
  {Kuznetsov}}, \bibinfo {author} {\bibfnamefont {A.~E.}\ \bibnamefont
  {Miroshnichenko}}, \bibinfo {author} {\bibfnamefont {M.~L.}\ \bibnamefont
  {Brongersma}}, \bibinfo {author} {\bibfnamefont {Y.~S.}\ \bibnamefont
  {Kivshar}}, \ and\ \bibinfo {author} {\bibfnamefont {B.}~\bibnamefont
  {Luk¡¯yanchuk}},\ }\href@noop {} {\bibfield  {journal} {\bibinfo  {journal}
  {Science}\ }\textbf {\bibinfo {volume} {354}},\ \bibinfo {pages} {aag2472}
  (\bibinfo {year} {2016})}\BibitemShut {NoStop}%
\bibitem [{\citenamefont {Yu}\ and\ \citenamefont {Capasso}(2014)}]{yu2014}%
  \BibitemOpen
  \bibfield  {author} {\bibinfo {author} {\bibfnamefont {N.}~\bibnamefont
  {Yu}}\ and\ \bibinfo {author} {\bibfnamefont {F.}~\bibnamefont {Capasso}},\
  }\href@noop {} {\bibfield  {journal} {\bibinfo  {journal} {Nature Mater.}\
  }\textbf {\bibinfo {volume} {13}},\ \bibinfo {pages} {139} (\bibinfo {year}
  {2014})}\BibitemShut {NoStop}%
\bibitem [{\citenamefont {Khorasaninejad}\ \emph {et~al.}(2016)\citenamefont
  {Khorasaninejad}, \citenamefont {Chen}, \citenamefont {Devlin}, \citenamefont
  {Oh}, \citenamefont {Zhu},\ and\ \citenamefont
  {Capasso}}]{khorasaninejad2016}%
  \BibitemOpen
  \bibfield  {author} {\bibinfo {author} {\bibfnamefont {M.}~\bibnamefont
  {Khorasaninejad}}, \bibinfo {author} {\bibfnamefont {W.~T.}\ \bibnamefont
  {Chen}}, \bibinfo {author} {\bibfnamefont {R.~C.}\ \bibnamefont {Devlin}},
  \bibinfo {author} {\bibfnamefont {J.}~\bibnamefont {Oh}}, \bibinfo {author}
  {\bibfnamefont {A.~Y.}\ \bibnamefont {Zhu}}, \ and\ \bibinfo {author}
  {\bibfnamefont {F.}~\bibnamefont {Capasso}},\ }\href@noop {} {\bibfield
  {journal} {\bibinfo  {journal} {Science}\ }\textbf {\bibinfo {volume}
  {352}},\ \bibinfo {pages} {1190} (\bibinfo {year} {2016})}\BibitemShut
  {NoStop}%
\bibitem [{\citenamefont {Wang}\ \emph
  {et~al.}(2018{\natexlab{c}})\citenamefont {Wang}, \citenamefont {Wu},
  \citenamefont {Su}, \citenamefont {Lai}, \citenamefont {Chen}, \citenamefont
  {Kuo}, \citenamefont {Chen}, \citenamefont {Chen}, \citenamefont {Huang},
  \citenamefont {Wang} \emph {et~al.}}]{wangshu2018}%
  \BibitemOpen
  \bibfield  {author} {\bibinfo {author} {\bibfnamefont {S.}~\bibnamefont
  {Wang}}, \bibinfo {author} {\bibfnamefont {P.~C.}\ \bibnamefont {Wu}},
  \bibinfo {author} {\bibfnamefont {V.-C.}\ \bibnamefont {Su}}, \bibinfo
  {author} {\bibfnamefont {Y.-C.}\ \bibnamefont {Lai}}, \bibinfo {author}
  {\bibfnamefont {M.-K.}\ \bibnamefont {Chen}}, \bibinfo {author}
  {\bibfnamefont {H.~Y.}\ \bibnamefont {Kuo}}, \bibinfo {author} {\bibfnamefont
  {B.~H.}\ \bibnamefont {Chen}}, \bibinfo {author} {\bibfnamefont {Y.~H.}\
  \bibnamefont {Chen}}, \bibinfo {author} {\bibfnamefont {T.-T.}\ \bibnamefont
  {Huang}}, \bibinfo {author} {\bibfnamefont {J.-H.}\ \bibnamefont {Wang}},
  \emph {et~al.},\ }\href@noop {} {\bibfield  {journal} {\bibinfo  {journal}
  {Nature Nanotech.}\ }\textbf {\bibinfo {volume} {13}},\ \bibinfo {pages}
  {227} (\bibinfo {year} {2018}{\natexlab{c}})}\BibitemShut {NoStop}%
\bibitem [{\citenamefont {Lin}\ \emph {et~al.}(2014)\citenamefont {Lin},
  \citenamefont {Fan}, \citenamefont {Hasman},\ and\ \citenamefont
  {Brongersma}}]{lin2014}%
  \BibitemOpen
  \bibfield  {author} {\bibinfo {author} {\bibfnamefont {D.}~\bibnamefont
  {Lin}}, \bibinfo {author} {\bibfnamefont {P.}~\bibnamefont {Fan}}, \bibinfo
  {author} {\bibfnamefont {E.}~\bibnamefont {Hasman}}, \ and\ \bibinfo {author}
  {\bibfnamefont {M.~L.}\ \bibnamefont {Brongersma}},\ }\href@noop {}
  {\bibfield  {journal} {\bibinfo  {journal} {science}\ }\textbf {\bibinfo
  {volume} {345}},\ \bibinfo {pages} {298} (\bibinfo {year}
  {2014})}\BibitemShut {NoStop}%
\bibitem [{\citenamefont {Decker}\ \emph {et~al.}(2015)\citenamefont {Decker},
  \citenamefont {Staude}, \citenamefont {Falkner}, \citenamefont {Dominguez},
  \citenamefont {Neshev}, \citenamefont {Brener}, \citenamefont {Pertsch},\
  and\ \citenamefont {Kivshar}}]{decker2015}%
  \BibitemOpen
  \bibfield  {author} {\bibinfo {author} {\bibfnamefont {M.}~\bibnamefont
  {Decker}}, \bibinfo {author} {\bibfnamefont {I.}~\bibnamefont {Staude}},
  \bibinfo {author} {\bibfnamefont {M.}~\bibnamefont {Falkner}}, \bibinfo
  {author} {\bibfnamefont {J.}~\bibnamefont {Dominguez}}, \bibinfo {author}
  {\bibfnamefont {D.~N.}\ \bibnamefont {Neshev}}, \bibinfo {author}
  {\bibfnamefont {I.}~\bibnamefont {Brener}}, \bibinfo {author} {\bibfnamefont
  {T.}~\bibnamefont {Pertsch}}, \ and\ \bibinfo {author} {\bibfnamefont
  {Y.~S.}\ \bibnamefont {Kivshar}},\ }\href@noop {} {\bibfield  {journal}
  {\bibinfo  {journal} {Adv. Opt. Mater.}\ }\textbf {\bibinfo {volume} {3}},\
  \bibinfo {pages} {813} (\bibinfo {year} {2015})}\BibitemShut {NoStop}%
\bibitem [{\citenamefont {Liu}(2017)}]{liu2017}%
  \BibitemOpen
  \bibfield  {author} {\bibinfo {author} {\bibfnamefont {W.}~\bibnamefont
  {Liu}},\ }\href@noop {} {\bibfield  {journal} {\bibinfo  {journal} {Phys.
  Rev. Lett.}\ }\textbf {\bibinfo {volume} {119}},\ \bibinfo {pages} {123902}
  (\bibinfo {year} {2017})}\BibitemShut {NoStop}%
\bibitem [{\citenamefont {Zhu}\ \emph {et~al.}(2015)\citenamefont {Zhu},
  \citenamefont {Kapraun}, \citenamefont {Ferrara},\ and\ \citenamefont
  {Chang-Hasnain}}]{zhu2015}%
  \BibitemOpen
  \bibfield  {author} {\bibinfo {author} {\bibfnamefont {L.}~\bibnamefont
  {Zhu}}, \bibinfo {author} {\bibfnamefont {J.}~\bibnamefont {Kapraun}},
  \bibinfo {author} {\bibfnamefont {J.}~\bibnamefont {Ferrara}}, \ and\
  \bibinfo {author} {\bibfnamefont {C.~J.}\ \bibnamefont {Chang-Hasnain}},\
  }\href@noop {} {\bibfield  {journal} {\bibinfo  {journal} {Optica}\ }\textbf
  {\bibinfo {volume} {2}},\ \bibinfo {pages} {255} (\bibinfo {year}
  {2015})}\BibitemShut {NoStop}%
\bibitem [{\citenamefont {Gholipour}\ \emph {et~al.}(2017)\citenamefont
  {Gholipour}, \citenamefont {Adamo}, \citenamefont {Cortecchia}, \citenamefont
  {Krishnamoorthy}, \citenamefont {Birowosuto}, \citenamefont {Zheludev},\ and\
  \citenamefont {Soci}}]{gholipour2017}%
  \BibitemOpen
  \bibfield  {author} {\bibinfo {author} {\bibfnamefont {B.}~\bibnamefont
  {Gholipour}}, \bibinfo {author} {\bibfnamefont {G.}~\bibnamefont {Adamo}},
  \bibinfo {author} {\bibfnamefont {D.}~\bibnamefont {Cortecchia}}, \bibinfo
  {author} {\bibfnamefont {H.~N.}\ \bibnamefont {Krishnamoorthy}}, \bibinfo
  {author} {\bibfnamefont {M.~D.}\ \bibnamefont {Birowosuto}}, \bibinfo
  {author} {\bibfnamefont {N.~I.}\ \bibnamefont {Zheludev}}, \ and\ \bibinfo
  {author} {\bibfnamefont {C.}~\bibnamefont {Soci}},\ }\href@noop {} {\bibfield
   {journal} {\bibinfo  {journal} {Adv. Mater.}\ }\textbf {\bibinfo {volume}
  {29}},\ \bibinfo {pages} {1604268} (\bibinfo {year} {2017})}\BibitemShut
  {NoStop}%
\bibitem [{\citenamefont {Geiss}\ \emph {et~al.}(2014)\citenamefont {Geiss},
  \citenamefont {Diziain}, \citenamefont {Steinert}, \citenamefont {Schrempel},
  \citenamefont {Kley}, \citenamefont {T{\"u}nnermann},\ and\ \citenamefont
  {Pertsch}}]{geiss2014}%
  \BibitemOpen
  \bibfield  {author} {\bibinfo {author} {\bibfnamefont {R.}~\bibnamefont
  {Geiss}}, \bibinfo {author} {\bibfnamefont {S.}~\bibnamefont {Diziain}},
  \bibinfo {author} {\bibfnamefont {M.}~\bibnamefont {Steinert}}, \bibinfo
  {author} {\bibfnamefont {F.}~\bibnamefont {Schrempel}}, \bibinfo {author}
  {\bibfnamefont {E.-B.}\ \bibnamefont {Kley}}, \bibinfo {author}
  {\bibfnamefont {A.}~\bibnamefont {T{\"u}nnermann}}, \ and\ \bibinfo {author}
  {\bibfnamefont {T.}~\bibnamefont {Pertsch}},\ }\href@noop {} {\bibfield
  {journal} {\bibinfo  {journal} {Phys. Status Solidi A}\ }\textbf {\bibinfo
  {volume} {211}},\ \bibinfo {pages} {2421} (\bibinfo {year}
  {2014})}\BibitemShut {NoStop}%
\bibitem [{\citenamefont {Chen}\ \emph {et~al.}(2011)\citenamefont {Chen},
  \citenamefont {Ng}, \citenamefont {Lin},\ and\ \citenamefont
  {Chan}}]{chen2011}%
  \BibitemOpen
  \bibfield  {author} {\bibinfo {author} {\bibfnamefont {J.}~\bibnamefont
  {Chen}}, \bibinfo {author} {\bibfnamefont {J.}~\bibnamefont {Ng}}, \bibinfo
  {author} {\bibfnamefont {Z.}~\bibnamefont {Lin}}, \ and\ \bibinfo {author}
  {\bibfnamefont {C.}~\bibnamefont {Chan}},\ }\href@noop {} {\bibfield
  {journal} {\bibinfo  {journal} {Nature Photo.}\ }\textbf {\bibinfo {volume}
  {5}},\ \bibinfo {pages} {531} (\bibinfo {year} {2011})}\BibitemShut {NoStop}%
\bibitem [{\citenamefont {Evlyukhin}\ \emph {et~al.}(2016)\citenamefont
  {Evlyukhin}, \citenamefont {Fischer}, \citenamefont {Reinhardt},\ and\
  \citenamefont {Chichkov}}]{evlyukhin2016}%
  \BibitemOpen
  \bibfield  {author} {\bibinfo {author} {\bibfnamefont {A.~B.}\ \bibnamefont
  {Evlyukhin}}, \bibinfo {author} {\bibfnamefont {T.}~\bibnamefont {Fischer}},
  \bibinfo {author} {\bibfnamefont {C.}~\bibnamefont {Reinhardt}}, \ and\
  \bibinfo {author} {\bibfnamefont {B.~N.}\ \bibnamefont {Chichkov}},\
  }\href@noop {} {\bibfield  {journal} {\bibinfo  {journal} {Phys. Rev. B}\
  }\textbf {\bibinfo {volume} {94}},\ \bibinfo {pages} {205434} (\bibinfo
  {year} {2016})}\BibitemShut {NoStop}%
\bibitem [{\citenamefont {Jackson}(2012)}]{jackson2012}%
  \BibitemOpen
  \bibfield  {author} {\bibinfo {author} {\bibfnamefont {J.~D.}\ \bibnamefont
  {Jackson}},\ }\href@noop {} {\emph {\bibinfo {title} {Classical
  electrodynamics}}}\ (\bibinfo  {publisher} {John Wiley \& Sons},\ \bibinfo
  {year} {2012})\BibitemShut {NoStop}%
\bibitem [{\citenamefont {Kravets}\ \emph {et~al.}(2018)\citenamefont
  {Kravets}, \citenamefont {Kabashin}, \citenamefont {Barnes},\ and\
  \citenamefont {Grigorenko}}]{kravets2018}%
  \BibitemOpen
  \bibfield  {author} {\bibinfo {author} {\bibfnamefont {V.}~\bibnamefont
  {Kravets}}, \bibinfo {author} {\bibfnamefont {A.}~\bibnamefont {Kabashin}},
  \bibinfo {author} {\bibfnamefont {W.}~\bibnamefont {Barnes}}, \ and\ \bibinfo
  {author} {\bibfnamefont {A.}~\bibnamefont {Grigorenko}},\ }\href@noop {}
  {\bibfield  {journal} {\bibinfo  {journal} {Chem. Rev.}\ } (\bibinfo {year}
  {2018})}\BibitemShut {NoStop}%
\bibitem [{\citenamefont {Kristensen}\ \emph {et~al.}(2017)\citenamefont
  {Kristensen}, \citenamefont {Yang}, \citenamefont {Bozhevolnyi},
  \citenamefont {Link}, \citenamefont {Nordlander}, \citenamefont {Halas},\
  and\ \citenamefont {Mortensen}}]{kristensen2017}%
  \BibitemOpen
  \bibfield  {author} {\bibinfo {author} {\bibfnamefont {A.}~\bibnamefont
  {Kristensen}}, \bibinfo {author} {\bibfnamefont {J.~K.}\ \bibnamefont
  {Yang}}, \bibinfo {author} {\bibfnamefont {S.~I.}\ \bibnamefont
  {Bozhevolnyi}}, \bibinfo {author} {\bibfnamefont {S.}~\bibnamefont {Link}},
  \bibinfo {author} {\bibfnamefont {P.}~\bibnamefont {Nordlander}}, \bibinfo
  {author} {\bibfnamefont {N.~J.}\ \bibnamefont {Halas}}, \ and\ \bibinfo
  {author} {\bibfnamefont {N.~A.}\ \bibnamefont {Mortensen}},\ }\href@noop {}
  {\bibfield  {journal} {\bibinfo  {journal} {Nature Rev. Mater.}\ }\textbf
  {\bibinfo {volume} {2}},\ \bibinfo {pages} {16088} (\bibinfo {year}
  {2017})}\BibitemShut {NoStop}%
\bibitem [{\citenamefont {Lee}\ \emph {et~al.}(2018)\citenamefont {Lee},
  \citenamefont {Jang}, \citenamefont {Jeong},\ and\ \citenamefont
  {Rho}}]{lee2018}%
  \BibitemOpen
  \bibfield  {author} {\bibinfo {author} {\bibfnamefont {T.}~\bibnamefont
  {Lee}}, \bibinfo {author} {\bibfnamefont {J.}~\bibnamefont {Jang}}, \bibinfo
  {author} {\bibfnamefont {H.}~\bibnamefont {Jeong}}, \ and\ \bibinfo {author}
  {\bibfnamefont {J.}~\bibnamefont {Rho}},\ }\href@noop {} {\bibfield
  {journal} {\bibinfo  {journal} {Nano Converg.}\ }\textbf {\bibinfo {volume}
  {5}},\ \bibinfo {pages} {1} (\bibinfo {year} {2018})}\BibitemShut {NoStop}%
\bibitem [{\citenamefont {Hu}\ \emph {et~al.}(2018)\citenamefont {Hu},
  \citenamefont {Lu}, \citenamefont {Cao}, \citenamefont {Zhang}, \citenamefont
  {Xu}, \citenamefont {Li}, \citenamefont {Gao}, \citenamefont {Cai},
  \citenamefont {Guan}, \citenamefont {Qiu} \emph {et~al.}}]{hu2018}%
  \BibitemOpen
  \bibfield  {author} {\bibinfo {author} {\bibfnamefont {D.}~\bibnamefont
  {Hu}}, \bibinfo {author} {\bibfnamefont {Y.}~\bibnamefont {Lu}}, \bibinfo
  {author} {\bibfnamefont {Y.}~\bibnamefont {Cao}}, \bibinfo {author}
  {\bibfnamefont {Y.}~\bibnamefont {Zhang}}, \bibinfo {author} {\bibfnamefont
  {Y.}~\bibnamefont {Xu}}, \bibinfo {author} {\bibfnamefont {W.}~\bibnamefont
  {Li}}, \bibinfo {author} {\bibfnamefont {F.}~\bibnamefont {Gao}}, \bibinfo
  {author} {\bibfnamefont {B.}~\bibnamefont {Cai}}, \bibinfo {author}
  {\bibfnamefont {B.-O.}\ \bibnamefont {Guan}}, \bibinfo {author}
  {\bibfnamefont {C.-W.}\ \bibnamefont {Qiu}},  \emph {et~al.},\ }\href@noop {}
  {\bibfield  {journal} {\bibinfo  {journal} {ACS Nano}\ }\textbf {\bibinfo
  {volume} {12}},\ \bibinfo {pages} {9233} (\bibinfo {year}
  {2018})}\BibitemShut {NoStop}%
\end{thebibliography}%


\section{Acknowledgements}
This work was supported by National Key R\&D Program of China (2017YFA0305100, 2017YFA0303800), National Natural Science Foundation of China (61775106, 11711530205, 11374006, 11774185, 91750204), the 111 Project (B07013), PCSIRT (IRT0149), and the Fundamental Research Funds for the Central Universities. We thank the Nanofabrication Platform of Nankai University for fabricating samples.

\end{document}